\documentclass[english,prl,twocolumn,showpacs,floatfix]{revtex4}
\usepackage{epsfig}
\usepackage{amssymb}
\usepackage{amsmath}
\usepackage{hyperref}
\usepackage{color}

\begin{document}
\newcommand{\kp}{$\mathbf{k}\!\cdot\!\mathbf{p}$}
\newcommand{\ldg}{electron Land{\'e} g-factor}
\newcommand{\dPfsT}{28.9}
\newcommand{\dPssT}{6.1}
\newcommand{\dPdT}{914}
\newcommand{\dPc}{-5.4}
\newcommand{\gstar}{-0.484}
\newcommand{\gstarpm}{0.7}

\newcommand{\E}[1]{\cdot 10^{#1}}
\newcommand{\U}[1]{\text{\rm #1}}
\newcommand{\mtc}[1]{\multicolumn{1}{c}{#1}}

\newcommand{\tcr}[1]{\textcolor{red}{#1}}
\newcommand{\hidden}[1]{}

\preprint{H{\"u}bner/PRL}

\title{{Temperature dependent Electron Land{\'e} g-Factor and Interband Matrix Element in
GaAs}}

\author{J. H{\"u}bner}
\email{jhuebner@nano.uni-hannover.de}
\author{S. D{\"o}hrmann}
\author{D. H{\"a}gele}\altaffiliation{Now at: Spectroscopy of Condensed Matter, Ruhr-Universit{\"a}t Bochum, 44801 Bochum, Germany}
\author{M. Oestreich}
\affiliation{Institute for Solid State Physics, Gottfried Wilhelm Leibniz
Universit{\"a}t Hannover, Appelstr. 2, 30167 Hannover, Germany}

\date{\today}

\begin{abstract}
Very high precision measurements of the electron Land{\'e} g-factor in GaAs are
presented using spin-quantum beat spectroscopy at low excitation densities and
temperatures ranging from 2.6 to 300\,K. In colligation with available data for
the temperature dependent effective mass a temperature dependence of the
interband matrix element within a common five level $\mathbf{k}
\!\cdot\!\mathbf{p}$ \,-\,theory can model both parameters consistently. A
strong decrease of the interband matrix element with increasing temperature
consistently closes a long lasting gap between experiment and theory and
substantially improves the modeling of both parameters.
\end{abstract}

\maketitle


The semiempirical \kp\,-\,theory is a universal tool to calculate the band
structure in semiconductors and semiconductor heterostructures and is regularly
employed in such different fields as the physics of semiconductor laser design,
the quantum Hall effect, and spintronics. The part of the theory describing
magnetic field related phenomena has been extensively improved since its
introduction by Kane \cite{Kane1957}, Luttinger and Kohn \cite{Luttinger1955}
in the mid fifties. Nowadays, 5- and more band \kp\,-\,models are state of the
art and many low temperature experiments have confirmed the incredible accuracy
of \kp\,-\,calculations \cite{Roth1959, Cardona1963, Hermann1977Hermann1984,
Hopkins1987, Pfeffer1990, Mayer1991}. All these experiments support the
validity of \kp\,-\,theory whereas a single but central experiment, which
measures the temperature dependence of the \ldg\ in GaAs, shows a strong
discrepancy between experiment and \kp\,-\,theory \cite{Oestreich1995}.

In this paper we present extremely high precision, temperature dependent
measurements of the \ldg\ and show that by introducing a temperature dependent
interband matrix element yields a consistent explanation for the temperature
dependence of the \ldg\ and the effective mass within common \kp\,-theory,
while keeping full temperature dependence on the very well known interband
critical points. The \kp\,-\,theory is a perturbation theory calculating the
electronic band-structure by expansion around a single point in the Brillouin
zone. In direct semiconductors like GaAs, the high symmetry $\Gamma$-point is
the natural expansion point. The only input parameters are in this case the
measured band-gaps at $\mathbf{k}=0$ and the interband matrix elements
($P,P',P'',\ldots $). The change of the band-gap energies with the lattice
temperature are very well known for GaAs and the only remaining relevant
parameter which does not possess a direct experimental access is the interband
matrix element $P$ or the related Kane energy $E_{P}=(2 m_{0}/\hbar^{2})P^{2}$,
respectively \cite{Kane1957}. The temperature dependence of $P$ has been
assumed to be marginal since $P$ is inversely proportional to the interatomic
distance $a$ \cite{Cardona1972} and the well known change of $a$ with
temperature $T$ due to anharmonic lattice potential is small. According to the
relation $E_{P}\propto 1/a^{2}$ between 0\,K and 300\,K $E_{P}$ should change
about $-0.4$\% or less \cite{Lawaetz1971,Shantharama1984}. However, this
procedure only considers the static change of $1/a^{2}$. With a
phenomenological approach of a temperature dependency of $E_{P}$ the
experimental data can only be correctly described by an about 14 times larger,
i.e., \dPc\% decrease of $E_{P}$ from 2.6\,K to room temperature.


The sample used in the experiment is bulk GaAs grown by molecular beam epitaxy
with a donor concentration of $1.2\cdot 10^{15}$\,cm$^{-3}$. The temperature
dependent \ldg\ is measured by spin-quantum-beat spectroscopy in the following
way: The sample is mounted in Voigt geometry in a split coil superconducting
magnet and excited with circular polarized light pulses from an 80\,MHz
picosecond laser. The sample temperature is varied from 2.6\,K to room
temperature, whereas the excess-energy of the exciting light is about 6\,meV
above the direct band gap for temperatures up to 80\,K. At higher temperatures,
the contribution of the excess energy is negligible compared to the thermal
energies present in the sample lattice. The excited carrier density is $6\cdot
10^{15}$\,cm$^{-3}$. The photoluminescence from the sample is collected in
backward direction. Energy- and time-resolution are performed by a spectrometer
followed by a synchroscan streak camera. The \ldg\ is deduced from the
oscillating time evolution of the cross circular polarized component of the
photoluminescence via the relation $g^{*}=\omega_{L} \hbar/(\mu_{B}B)$, with
$\omega_{L}$ being the Larmor precession frequency of the conduction electron
spins, $\mu_{B}$ Bohr's magneton, and $B$ the magnetic field.

Great care has been taken in the time calibration of the detection system as
well as in the correct determination of the magnetic field present in the
superconducting magnet. Superconducting magnets can often show unapparent
remanence fields and incorrect field calibrations which easily influence the
experimental data. Therefore the magnetic field is calibrated with a precise
Hall sensor for all applied fields. The often cited value of $g^{*}=-0.44(2)$
\cite{Weisbuch1977} in GaAs applies for donor bound electrons, whereas for the
free conduction band electrons higher absolute g-factor values are reported
\cite{Krapf1990, Schreiner1997}. From the data presented here, a very high
accuracy extrapolated \ldg\ of $g^{*}=\gstar(3)$ at $T=0$\,K, $B=0$\,T, and
$P_{\rm exc.}=0$\,mW is determined.

\begin{figure}
\epsfig{file=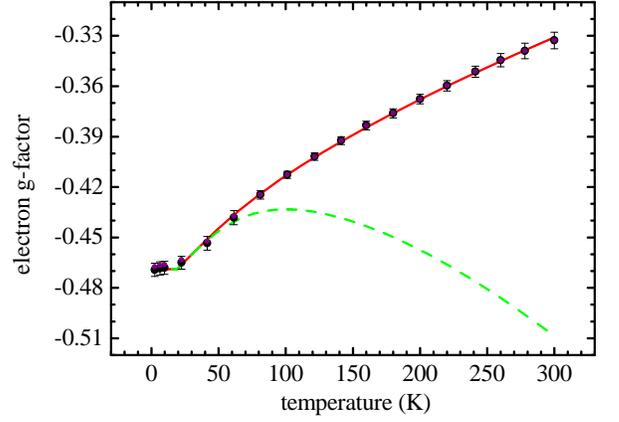,width=0.9\columnwidth} %
\caption{(Color online) High precision measurement of the temperature
dependence of the \ldg\ in bulk GaAs (filled circles).  The red solid line is a
fit of the experimental data by Eq. \ref{eq:gstar} with a strong temperature
dependent interband matrix element. The green dashed line is the calculated
$g^{*}$ with $E_{P}(T)$ depending only on the anharmonic lattice expansion.
Both calculations contain both the dilatational and electron-phonon
contribution to the change in the interband critical points according to
Eqn.~\ref{eq:EfromT} and values shown in Tab.~\ref{tab:parameters}.
}\label{fig:gstar}
\end{figure}

Figure \ref{fig:gstar} shows the \ldg\ versus sample temperature. Each value is
extrapolated to zero magnetic field and zero excitation power from measurements
at different fields \cite{BField} and excitation powers at constant temperature
to eliminate any residual effects of those entities on the g-factor. The
measurements from 2.6 to 62\,K are carried out with alternating excitation of
left and right circular polarized light and a small tilt of the sample against
the magnetic field. This technique enables us to monitor and subtract the
influence of the effective nuclear field on the \ldg\ measurement (see Ref.
\cite{Dohrmann2008} for details). This procedure is especially at low
temperatures much more precise concerning the absolute value of $g^{*}$ than
compared with other techniques like e.g. Ref. \cite{Hohage2006}.

Next the experimental results are compared with established 5-level
\kp\,-\,theory \cite{higherorder}. Cardona was the first to suggest a 5-level
approach based on the wavefunction expansion of the isoelectric group IV
counterpart \cite{Cardona1963}. However, the central 5-level \kp\,-\,result for
$g^{*}$ and $m^{*}$ used in this work was put forward by Hermann \textit{et
al}. \cite{Hermann1977Hermann1984} with $P$ and $P'$ as free parameters:
{
\begin{eqnarray}
\frac{{g^* }}{{g_0 }} &=& 1 - \frac{{E_P }}{3}\left( {\frac{1}{{E(\Gamma _6^c -
\Gamma _8^v )}} - \frac{1}{{E(\Gamma _6^c  - \Gamma _7^v )}}} \right) \label{eq:gstar} \\ 
&-& \frac{{E_{P'} }}{3} \left( {\frac{1}{{E(\Gamma _7^c  - \Gamma _6^c )}} -
\frac{1}{{E(\Gamma _8^c  - \Gamma _6^c )}}} \right) + \Delta^{g}_{\text{so}} +
C'\quad
\nonumber\\
\frac{{m_0 }}{{m^* }} &=& 1 + \frac{{E_P }}{3}\left( {\frac{2}{{E(\Gamma _6^c -
\Gamma _8^v )}} + \frac{1}{{E(\Gamma _6^c  - \Gamma _7^v )}}} \right) \label{eq:mstar} \\ 
&-& \frac{{E_{P'} }}{3} \left( {\frac{1}{{E(\Gamma _7^c  - \Gamma _6^c )}} +
\frac{2}{{E(\Gamma _8^c  - \Gamma _6^c )}}} \right) + \Delta^{m}_{\text{so}} +
C \nonumber
\end{eqnarray}
}%
Here, $g_{0}=2.0023$ is the free \ldg, $m_{0}$ the free electron mass, and
$E(\Gamma_{7}^{v}, \Gamma_{8}^{v}, \Gamma_{6}^{c}, \Gamma_{7}^{c},
\Gamma_{8}^{c})$ are the energies of the band extrema at the center of the
Brillouin zone. The correction $\Delta^{g}_{\text{so}}$ due to the spin orbit
coupling $\bar \Delta =-50$\,meV \cite{Sigg1987} between the valence and higher
conduction band has a significantly strong contribution to the
\ldg\ \cite{Pfeffer1990,Cardona1988} and is given by: {
\begin{equation}
\begin{split}
\Delta^{g}_{\text{so}} =  - \frac{2}{9} \bar \Delta \sqrt {E_P E_{P'} } \biggl(
\frac{2}{{E(\Gamma _6^c  - \Gamma _7^v ) \cdot E(\Gamma _7^c - \Gamma _6^v )}}
+ \\
\frac{1}{{E(\Gamma _6^c  - \Gamma _8^v ) \cdot E(\Gamma _8^c  - \Gamma _6^v )}}
\biggr)
\end{split}
\end{equation}
}%
Further corrections by $\bar\Delta$ to the terms linear in $E_{P,P'}$ in
Eqs.~\ref{eq:gstar} and \ref{eq:mstar} change $g^{*}$ below 0.1\% and are
neglected. The corresponding correction for the effective mass
$\Delta^{m}_{\text{so}}$ is included in the calculation as well, but has a much
smaller effect on the calculation of $m^{*}$ than $\Delta^{g}_{\text{so}}$ has
on $g^{*}$. The contributions from higher bands are summarized in the constants
$C'=-0.021$ and $C=-1.878$ taken from Ref. \cite{Mayer1991}.

The spin orbit interaction results mainly from contributions of the atomic
species in the material which is in good approximation temperature independent.
The temperature dependence of the band-gap energies $E(T)$ are very well known
for GaAs by experiment and described by the semi-phenomenological model
introduced by Vi\~na et al. \cite{Vina1984} collecting the electron and phonon
dynamics in one context:
\begin{equation}\label{eq:EfromT}
  E(T)=E_{B}-\alpha_{B}\left(1+\frac{2}{e^{\Theta/T}-1}\right).
\end{equation}
This model is used for all following calculations, but nearly identical results
are obtained by using the popular empirical relation by Varshni
\cite{Varshni1967}. The parameters for the \kp\,-\,calculations are listed in
Tab. \ref{tab:parameters}.

Equation \ref{eq:gstar} yields the \ldg\ at the conduction band minimum. At
finite temperatures, the known energy dependence of the \ldg\ $g^*(E)=g^* +
6.3\,\U{eV}^{-1}\cdot E$ on the kinetic energy $E$ of the electrons in GaAs
(Ref. \cite{Hopkins1987}, Fig. 7) is included by weighting $g^*$ with the
thermal distribution \cite{LandauLevels} of the electrons in the conduction
bands:
\begin{equation}\label{eq:tdep_g} \langle
g^{*}\rangle=\frac{\int_{0}^{\infty}dE\, g^{*}(E)\,
D^{3\rm D}(E)\,e^{\frac{-E}{k_{\rm B}T_{\rm e}}}}%
{\int_{0}^{\infty}dE\, D^{3\rm D}(E)\,e^{\frac{-E}{k_{\rm B}T_{\rm e}}}}.
\end{equation}
The integration starts at the minimum of the conduction band, $D^{3\rm
D}(E)=\,^{1}\!/_{2\pi^{2}}\left(^{2\,m^{*}}\!\!/_{\hbar^{2}}\right)^{3/2}
\sqrt{E}$ is the three-dimensional density of states, $k_{\rm B}$
Boltzmann's constant, and $T_{\rm e}$ the effective electron temperature.
The temperature dependence of the effective conduction band mass
\begin{table}
\caption{\label{tab:parameters}Parameters used}
\begin{ruledtabular}
\begin{tabular}{lllll}
                              & \mtc{$E_{B}$}& \mtc{$\alpha_{B}$}  & \mtc{$\Theta$} & \\
                              & \mtc{eV}    & \mtc{meV}           & \mtc{K}        & \mtc{Ref.}\\\hline
$E(\Gamma_{6}^{c}-\Gamma_{8}^{v})$           & 1.571\,eV   & 57                  & 240            & \footnotemark[1]\\
$E(\Gamma_{6}^{c}-\Gamma_{7}^{v})$  & 1.907\,eV   & 58                  & 240            & \footnotemark[1],\footnotemark[2]\\
$E(\Gamma_{7}^{c}-\Gamma_{8}^{v})$           & 4.563\,eV   & 59                  & 323            & \footnotemark[1]\\
$E(\Gamma_{8}^{c}-\Gamma_{8}^{v})$           & 4.718\,eV   & 59                  & 323            & \footnotemark[3],\footnotemark[4]\\
$E_{P}$                       & \dPfsT\,eV   & \dPdT               & 240            & \footnotemark[5],\footnotemark[6]\\
$E_{P'}$                      & \dPssT\,eV    & \dPdT               & 240            & \footnotemark[5],\footnotemark[6],\footnotemark[7]\\
\end{tabular}
\end{ruledtabular}
\footnotetext[1]{from Ref. \cite{Lautenschlager1987}.} %
\footnotetext[2]{The value for $\alpha$ for the Varshni model in
\cite{Lautenschlager1987} contains a typing error. It should read
$\alpha=5.4\cdot 10^{-4}\,\U{eV/K}$ as seen from the data presented.}
\footnotetext[3]{from Ref. \cite{Hermann1977Hermann1984}. Here
$\alpha_{B}=59$\,meV has been added to $E_{B}$
to account for the Vi\~na model used in this work.} %
\footnotetext[4]{Assumption that the temperature dependence of the
$E(\Gamma_{8}^{c})$ is the same as for the $E(\Gamma_{7}^{c})$ band due to
the lack of available data. The induced error is small, since these values
contribute only weakly to the g-factor
correction.} %
\footnotetext[5]{Given by the value of $g^{*}$ and $m^{*}$ at T=0\,K.}
\footnotetext[6]{The ratio of $E_{P}$ and $E_{P'}$ compares to that estimated
by Cardona \textit{et al.} \cite{Cardona1963,Cardona1988}. Note that $E_{P}(T)
[\text{[eV]}] =(2 m_{0}/\hbar^{2})(P^{(')}[\text{eV m}])^{2}(T)$.
}%
\footnotetext[7]{Assumption that the temperature dependence of $E_{P}$ is the
same as for the $E_{P'}$. See also \footnotemark[4].}
\end{table}
$m^{*}$ has been taken into account in $D^{3\rm D}$ according to
Eq.~\ref{eq:mstar}. At lattice temperatures $T_{\rm L}$ below 20\,K, the
effective electron temperature $T_{\rm e}$ is constant due to the excess energy
of the optical excitation and phase space filling. At higher lattice
temperatures, the electron-phonon coupling is much more efficient and $T_{\rm
e}$ is in good approximation equal to the lattice temperature, and phase space
filling can be neglected for the calculation of $ \langle g^{*}\rangle$ due to
the low excitation densities.

The green dashed line in Fig.~\ref{fig:gstar} shows $\langle g^{*}\rangle$
calculated with Eq.~\ref{eq:tdep_g} including the weakly temperature dependent
interband matrix element due to the anharmonic lattice potential alone evincing
that the model is in clear disagreement with the measurements. In the next
step, additionally the same temperature relation for the Kane energies, i.e.,
the interband matrix elements is assumed as for the band-gap energies in
Eq.~\ref{eq:EfromT} and the linear pre-factor $\alpha_{B}$ is the \emph{only}
fit parameter, keeping $\Theta$ fixed to 240\,K. The resulting red solid line
in Fig.~\ref{fig:gstar} exhibits excellent agreement with the measurement. This
fit however implies that the interband matrix element reduces from helium to
room temperature by as much as \dPc\% within this model and can not be
explained by the tiny average lattice expansion expected from the anharmonic
lattice potential alone.

\begin{figure}
\epsfig{file=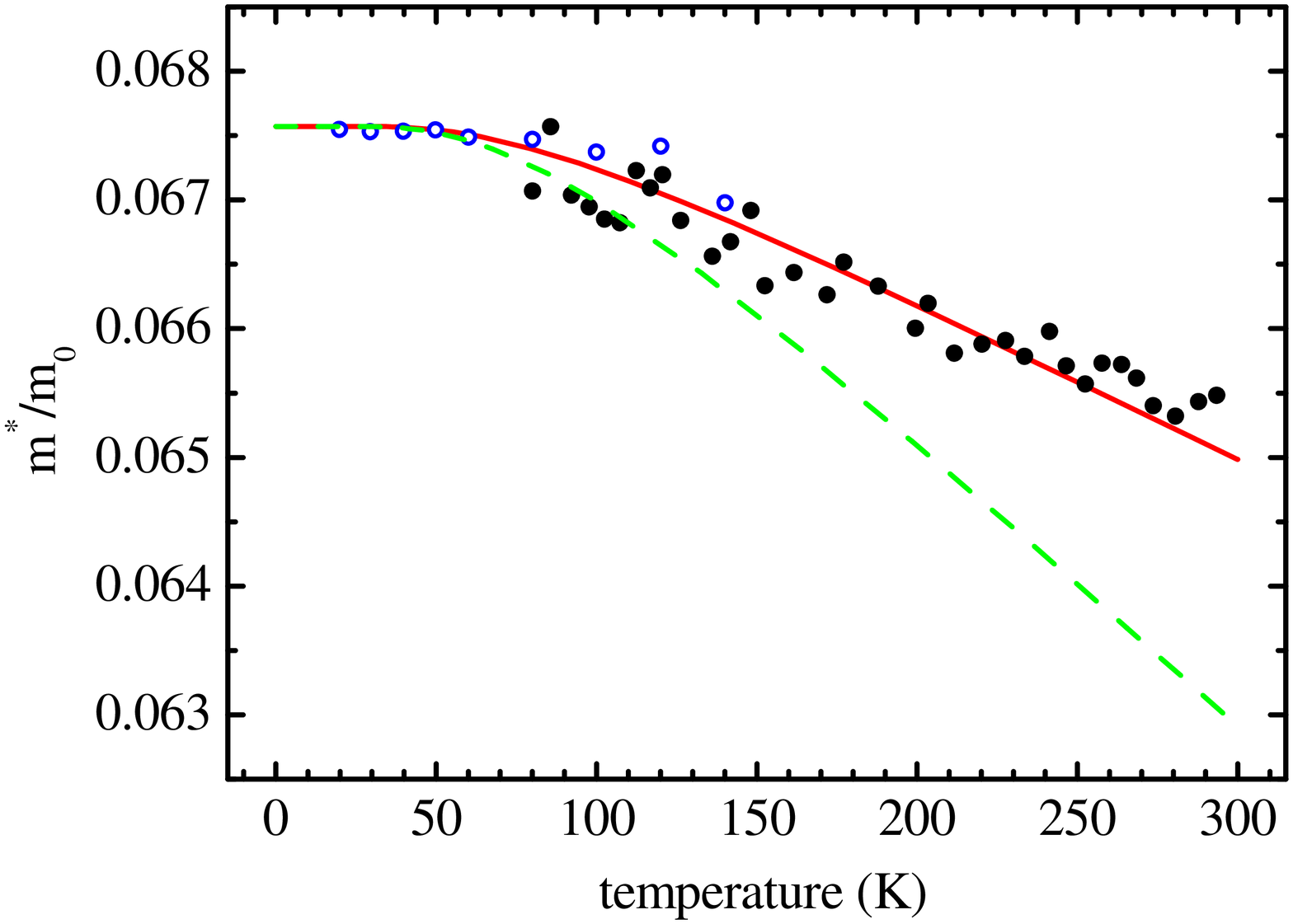,width=0.9\columnwidth} %
\caption{(Color online) Temperature dependence of the effective conduction
electron mass in bulk GaAs (hollow blue circles \cite{Hopkins1987}, full black
circles \cite{Hazama1986}). The red solid line follows Eq. \ref{eq:gstar} with
a strong temperature dependent interband matrix. The green dashed line is
calculated with the conventional temperature dependence of $E_{P}$. Please note
the different impact of $E_{P}$ on $g^{*}$ and $m^{*}$ acc. to Eq.
\ref{eq:gstar} and \ref{eq:mstar}.}\label{fig:mstar}
\end{figure}

To substantiate the possibility of a strong temperature dependence
of the interband matrix elements, the experimentally determined
temperature dependence of the effective mass in GaAs is compared
with predictions by the same 5-level \kp\,-\,model.
Figure~\ref{fig:mstar} shows the temperature dependent effective
mass of GaAs measured by cyclotron resonance \cite{Hopkins1987}
and magneto-phonon \cite{Hazama1986} spectroscopy. The data
presented from these publications represent the 'bare' effective
mass at the conduction band minimum (cf. \cite{Stradling1968}).
The red solid line in Fig. \ref{fig:mstar} depicts the calculated
temperature dependence of $m^*$ (Eq.~\ref{eq:mstar}) including the
strong dependency of the interband matrix elements on the
temperature according to Eq. \ref{eq:EfromT}. Only parameters
consistently obtained with the $g^*$-data according to Tab.
\ref{tab:parameters} are employed, i.e., the calculation of
$m^{*}$ has \emph{no} free parameter. Nevertheless the calculation
is in excellent agreement with the experiment, whereas the
discrepancy between conventional \kp\,-\,theory and experiment is
obvious: The green dashed line in Fig. \ref{fig:mstar} shows the
same calculation but with the old established temperature
dependence of the interband matrix elements.

We want to illustrate a semiclassical possible physical origin of a stronger
temperature dependence of $P$. The presence of longitudinal acoustical phonons
creates a locally and temporally varying conduction band energy leading to
electronic states with lowered energy via the deformation potential distributed
over a range of phonon wavelengths corresponding to the occupancy of the phonon
energies. Similarly, hole states with higher energies appear in the band
structure. Note that the fraction of the well known band gap shrinkage due to
electron-phonon interaction with elevated temperatures is treated differently
\cite{Fan1951,Allen1976}. For phonon wavelengths longer than the electron
scattering length the free conduction band electrons are fast compared to the
lattice dynamics and can follow the local, phonon-induced conduction band
minima adiabatically on small scales. As a consequence the electrons average
preferentially over elongated lattice sites and the wavefunction overlap
between conduction and valence band states, i.e., the interband matrix element
is reduced. The mean square relative displacement for all occupied longitudinal
acoustic phonon branches is easily calculated and yields a reduction of $E_{P}$
by -0.4\,\% at 100\,K and -1.6\,\% at 300\,K compared to 0\,K. This estimated
shrinkage is already a factor of four bigger than the established shrinkage due
to the anharmonicity of the lattice potential. Modeling \hidden{AE} the
temperature dependence of $g^*$ and $m^*$ can be as well pursued by inserting
only the dilatational part of the interband critical points instead of taking
into account the full temperature dependence of the interband critical points
and matrix elements. Due to the smaller contribution of the dilatational part
to the interband critical points, their change is less strong than the full
change in the optical gap and the result is mathematically nearly identical
compared to the approach following Eqns.~\ref{eq:gstar} and \ref{eq:mstar}. The
concept of the effective mass band gap $E_{g}^{*}$ \cite{Ehrenreich1957} would
accordingly be extended to the ``effective \ldg\ band gap" in this work.
However, both concepts cannot be distinguished within the same \kp-description
employed here and the full temperature behavior of the interband critical
points is much more precisely known, than their change with lattice constant,
i.e their pressure dependence. Nevertheless both effects can yield a
contribution at the same time, especially since electron-phonon interaction has
a small but non vanishing contribution affecting the band curvature
\cite{Froehlich1950}. The contemplations involving the dynamical change in
interatomic distances like those pointed out above for $E_{P}$ should be
included in future analyses to improve on the modeling of the central
quantities of the employed 5-level \kp-model.



In summary, the \ldg\ in GaAs has been determined with very high precision in
dependence on the sample temperature resulting in a free conduction band
g-factor of $\gstar(3)$ at T=0\,K. The experimental data on the temperature
dependence of $g^{*}$  and $m^{*}$ have been consistently modeled \hidden{AE}
with a modified \kp\,-\,formalism. We suggest to include phonon induced lattice
fluctuations similar to the band gap shrinkage of semiconductors, which support
the experimental findings.

\begin{acknowledgments}
The authors thank Roland Winkler for helpful discussions and appreciate H.J.
Queisser's inspiration. The work is founded by the German Science Foundation
(DFG - Priority Program 1285 "Semiconductor Spintronics"), the Federal Ministry
for Education and Research (BMBF - Program NanoQUIT) and QUEST Hannover.
\end{acknowledgments}


\end{document}